\documentclass[letterpaper,twocolumn,10pt]{article}
\usepackage{usenix2019_v3}

\usepackage{tikz}
\usepackage{amsmath}
\usepackage{tabularx}

\begin{document}

\date{}

\title{\Large \bf Intrinsic Propensity for Vulnerability in Computers? \\ Arbitrary Code Execution in the Universal Turing Machine}

\author{
Pontus Johnson\\
KTH Royal Institute of Technology\\
Stockholm, Sweden\\
pontusj@kth.se\\

} 

\maketitle

\begin{abstract}
The universal Turing machine is generally considered to be the simplest, most abstract model of a computer. This paper reports on the discovery of an accidental arbitrary code execution vulnerability in Marvin Minsky's 1967 implementation of the universal Turing machine. By submitting crafted data, the machine may be coerced into executing user-provided code. The article presents the discovered vulnerability in detail and discusses its potential implications. To the best of our knowledge, an arbitrary code execution vulnerability has not previously been reported for such a simple system.
\end{abstract}

\section{Introduction}
\label{intro}

Arbitrary code execution holds a special position among malicious exploits. Most remarkable is the case when such code execution is effected through the submission of crafted data. In computing, the relationship between structure and behavior, between program and process, is perplexing in itself. That this relationship so often can be subverted, allowing an untrusted data provider to preternaturally gain control over program execution, is disquieting. Why is this a common phenomenon in computer systems? Is it the consequence of incidental but unfortunate decisions in the development history of those systems, or is it rather the result of some fundamental property of computing? 

Commonly used to explore the foundational traits of computers and computing, the \textit{universal Turing machine} is generally considered one of the most important ideas in computer science. Turing presented his universal machine in a paper in 1936 \cite{turing1936computable}, where he promptly used it to solve one of the most pressing mathematical questions of the day, David Hilbert and Wilhelm Ackermann so called \textit{Entscheidungsproblem} \cite{hilbert1938grundzuge}. As expressed by Marvin Misky, ``the universal machine quickly leads to some striking theorems bearing on what appears to be the ultimate futility of attempting to obtain effective criteria for effectiveness itself'' \cite{minsky1967computation} . But the universal Turing machine achieved more than that. As stated by Davis, Sigal and Weyuker in \cite{davis1994computability}, ``Turing's construction of a universal computer in 1936 provided reason to believe that, at least in principle, an all-purpose computer would be possible, and was thus an anticipation of the modern digital computer.'' Or, in the words of Stephen Wolfram \cite{wolfram2002new}, `what launched the whole computer revolution is the remarkable fact that universal systems with fixed underlying rules can be built that can in effect perform any possible computation.'' Not only the universality, but also the simplicity of the universal Turing machine has attracted interest. In 1956, Claude Shannon explored some minimal forms of the universal Turing machine \cite{shannon1956universal}, and posed the challenge to find even smaller such machines. That exploration has continued to this day \cite{wolfram2002new}.

A common strategy for understanding a problem is to reduce it to its minimal form. In the field of computer security, we may ask the question: "What is the simplest system exploitable to arbitrary code execution?" In this article, we propose an answer to that question by reporting on the discovery that a well-established implementation \cite{minsky1967computation} of the universal Turing machine is vulnerable to a both unintentional and non-trivial form of arbitrary code execution. 

The article proceeds in the next section with a background to arbitrary code execution. Section \ref{utm} reviews universal Turing machines, and in particular the studied implementation. This is followed by a detailed analysis of the discovered vulnerability. In Section \ref{mitigations}, we consider changes to the explored implementation that would mitigate the vulnerability. The paper concludes with a discussion on the significance of the findings, and some conclusions. 

\section{Arbitrary Code Execution}
\label{ace}

That a software user who is nominally only granted the possibility to provide some trivial data, such as her name, sometimes, by carefully crafting that seemingly inconsequential data, is able to take full control of the computer executing that software, is remarkable indeed. It is even more arresting that such arbitrary code execution vulnerabilities are quite frequently discovered in software systems. Arbitrary code execution is not a fringe phenomenon, but a material class of vulnerabilities in modern computer systems. There are several specific types of vulnerabilities that may lead to arbitrary code execution. Among the 2019 CWE (Common Weakness Enumeration) Top 25 Most Dangerous Software Errors the following may lead to arbitrary code execution \cite{mitre2019cwe}, 

\begin{table}[!htbp]
  \centering
    \begin{tabularx}{\columnwidth}{|l|X|}
CWE-119 & Improper Restriction of Operations within the Bounds of a Memory Buffer \\
CWE-79 & Improper Neutralization of Input During Web Page Generation \\
CWE-20 & Improper Input Validation \\
CWE-89 & Improper Neutralization of Special Elements used in an SQL Command \\
CWE-416 & Use After Free \\
CWE-190 & Integer Overflow or Wraparound \\
CWE-78 & Improper Neutralization of Special Elements used in an OS Command \\
CWE-787 & Out-of-bounds Write \\
CWE-476 & NULL Pointer Dereference \\
CWE-434 & Unrestricted Upload of File with Dangerous Type \\
CWE-94 & Improper Control of Generation of Code \\
CWE-502 & Deserialization of Untrusted Data \\
    \end{tabularx}%
  \label{tab:addlabel}%
  \caption{Vulnerabilities that may lead to code execution}
\end{table}%

Those twelve constitute half of that top 25 list, highlighting the prevalence of this class of vulnerability. It is not, however, clear whether there is any common underlying cause to these vulnerabilities; is there any root explanation as to why they are so prevalent? 

\section{Universal Turing machine}
\label{utm}
As preparation for the presentation of the arbitrary code execution vulnerability in Section \ref{exploit}, we review the concept of the Turing machine, the universal Turing machine, and the Minsky implementation of that universal machine. 

\subsection{Turing machine}
\begin{figure}[!t]
\centering
\includegraphics[width=0.5\textwidth,keepaspectratio]{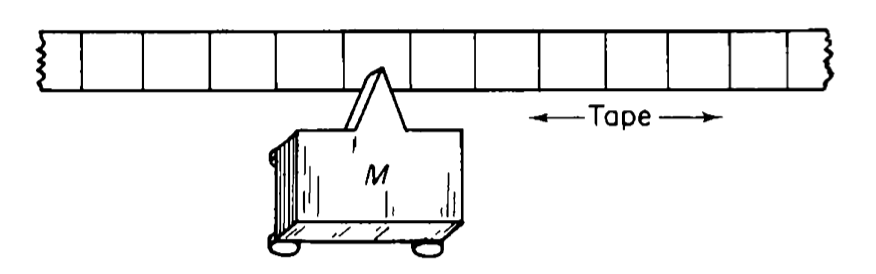}
\caption{A Turing machine. \label{fig-tm}}
\end{figure}

A \textit{Turing machine}, $T$, is a \textit{finite-state machine} operating on a \textit{tape} by means of the machine's \textit{head} (cf. Figure \ref{fig-tm}). The tape has the form of a sequence of squares onto one of which the head is positioned. The head can read and write symbols located in the currently scanned square. It can also move one square to the left or right. 

The input to the finite-state machine is the currently scanned symbol, while the output is the printed symbol as well as the direction in which the head is to move. The finite-state machine, which thus controls the actions of the head, can therefore be represented as a quintuple, 
$$
Q_iS_i\ Q_{ij}S_{ij}D_{ij}
$$

where $Q_i$ represents the source state, $S_i$ the scanned symbol, $Q_{ij}$ the target state, $S_{ij}$ the printed symbol and $D_{ij}$ the direction in which the head is to move.

\subsection{Universal Turing machine}
A \textit{universal} Turing machine, $U$, is a Turing machine that is capable of simulating any other Turing machine, $T$. There are multiple implementations of the universal Turing machine, the first one notably being the one proposed by Alan Turing himself in \cite{turing1936computable}. In this article, we consider the universal Turing machine proposed by Marvin Minsky in \cite{minsky1967computation} . Our choice of the Minsky version is mainly based on (i) the ease with which it can be implemented, (ii) the ease with which it can be explained in a brief article, and (iii) it's solid place in computer science literature, presented by Marvin Minsky in his much-cited book \textit{Computation: Finite and Infinite Machines} (1967). Turing's own universal machine is arguably more convoluted, and also contains a number of errors \cite{copeland2004essential}. Other universal Turing machines include a set of minimally small universal Turing machines, counting the size of their alphabet and finite-machine state space \cite{rogozhin1996small}\cite{woods2009complexity}. Those machines, however, add cognitive complexity by introducing an additional formalism (a \textit{tag system}) in order to minimize the size of the machines. 

\begin{figure}[!t]
\centering
\includegraphics[width=0.5\textwidth,keepaspectratio]{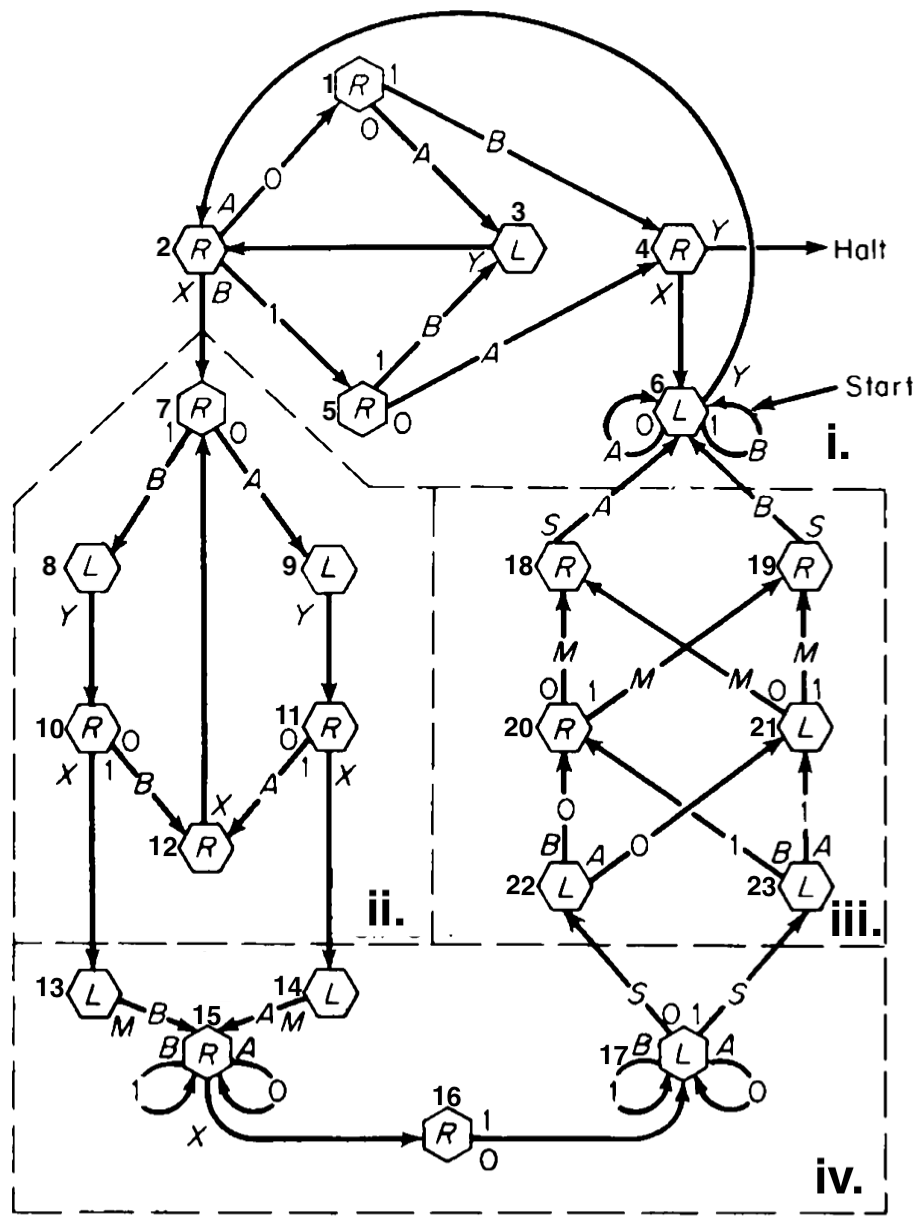}
\caption{Finite-state machine of Marvin Minsky's universal Turing machine. \label{fig-utm}}
\end{figure}

\subsubsection{Machine structure}
In the words of Minsky himself, the universal machine, $U$, 

\begin{quotation}
will be given just the necessary materials: a description, on its tape, of $T$ and of [the initial configuration on $T$s own, simulated tape] $s_x$; some working space; and the built-in capacity to interpret correctly the rules of operation as given in the description of $T$. Its behavior will be very simple. $U$ will simulate the behavior of $T$ one step at a time. It will be told by a marker \texttt{M} at what point on its tape $T$ begins, and then it will keep a complete account of what $T$'s tape looks like at each moment. It will remember what state $T$ is supposed to be in, and it can see what $T$ would read on the 'simulated' tape. Then $U$ will simply look at the description of $T$ to see what $T$ is next supposed to do, and do it! This really involves no more than looking up, in a table of quintuples, to find out what symbol to write, which way to move, and what new state to go into. We will assume that $T$ has a tape which is infinite only to the left, and that it is a binary (2-symbol) machine. These restrictions are inessential, but make matters much simpler.
 \end{quotation}
 
Concretely, $U$s tape is divided into four regions. The infinite region to the left will be the tape of the simulated machine $T$. The second region, $q(t)$, contains the name of the current state of $T$. The third region, $s(t)$ stores the value of the symbol under $T$'s head. Together, we denote $q(t)s(t)$ the \textit{machine condition}. The fourth region, $d_T$ will contain the machine description of $T$, i.e. the program. 

\begin{figure}[!t]
\centering
\includegraphics[width=0.5\textwidth,keepaspectratio]{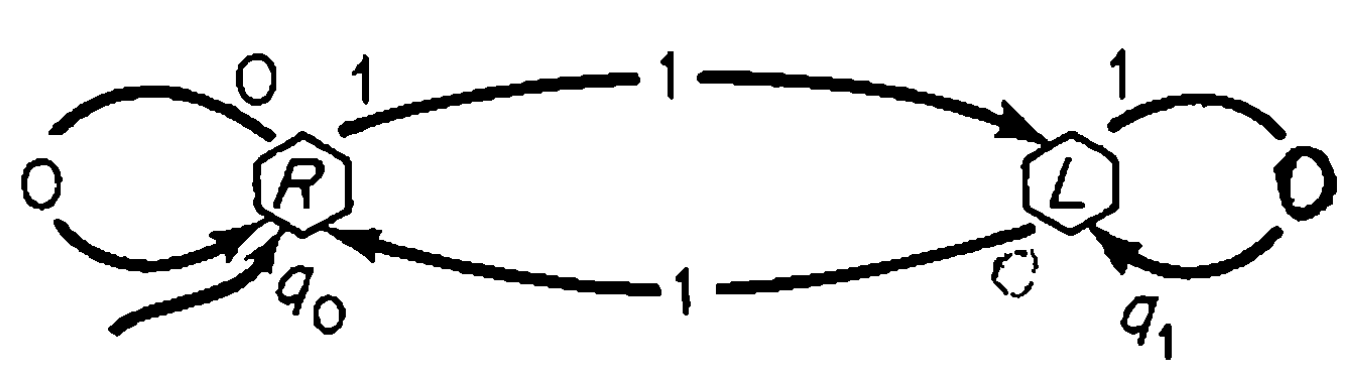}
\caption{The state machine of a binary counter Turing machine. \label{fig-bc}}
\end{figure}

If $U$ is to simulate the binary counter represented in Figure \ref{fig-bc}, $U$'s tape may initially be configured as follows,

\begin{flushleft}
\begin{footnotesize}
\mbox{\texttt{...00000M000Y001X0000001X0010110X0100011X0110100Y00...}} \newline
\mbox{\texttt{\ \ \ \ \ \ \ \ \ \ \ \ \ \ \ \ $\uparrow$}}
\end{footnotesize}
\end{flushleft}

where the arrow points to the location of $U$'s head, the \texttt{M} marks the location of $T$'s head, the leftmost \texttt{Y} separates $T$'s tape from $q(t)$, and the leftmost \texttt{X} identifies the start of the machine description, $d_T$. Within $d_T$, quintuples are separated by \texttt{X}s, and the rightmost \texttt{Y} marks the end of the machine description.

The machine description of $T$ is constituted of a set of quintuples, $Q_iS_iQ_{ij}S_{ij}D_{ij}$, recorded in binary format. An example would be \texttt{0010110}, where $Q_i=\mathtt{00}$, $S_i=\mathtt{1}$, $Q_{ij}=\mathtt{01}$, $S_{ij}=\mathtt{1}$, and $D_{ij}=\mathtt{0}$. Any number of binary digits may be used to represent the machine state, $Q$. We adopt the convention that $D_{ij}=\mathtt{0}$ indicates a shift of the head to the left, while $D_{ij}=\mathtt{1}$ shifts the head to the right.

\subsubsection{Machine execution}
$U$'s finite-state machine, presented in Figure \ref{fig-utm}, is constituted of four distinct phases, marked as i-iv in the figure. The first phase uses $T$'s state, $q(t)$, and the symbol under $T$'s head, $s(t)$, to identify the next quintuple to execute. A recurring approach for marking positions is to recast \texttt{0}s and \texttt{1}s into \texttt{A}s and \texttt{B}s. The example tape presented above would by the first phase be modified to

\begin{flushleft}
\begin{footnotesize}
\mbox{\texttt{...00000M000Y001XAAAAAABXAAB0110X0100011X0110100Y00...}} \newline
\mbox{\texttt{\ \ \ \ \ \ \ \ \ \ \ \ \ \ \ \ \ $\uparrow$}}
\end{footnotesize}
\end{flushleft}

where the transition from  \texttt{A}s and \texttt{B}s into \texttt{0}s and \texttt{1}s specifies the position of the quintuple (\texttt{0010110}, in the example) matching $q(t)$.

From the identified quintuple, the second phase copies the target state, $Q_{ij}$ (\texttt{01}), and the symbol to be written, $S_{ij}$ (\texttt{1}), to $q(t)$ and $s(t)$, and remembers the direction $D_{ij}$ by entering into the appropriate state, (state 13 or 14 in Figure \ref{fig-utm}).

\begin{flushleft}
\begin{footnotesize}
\mbox{\texttt{...00000M000YABBXAAAAAABXAABABBAX0100011X0110100Y00...}} \newline
\mbox{\texttt{\ \ \ \ \ \ \ \ \ \ \ \ \ \ \ \ $\uparrow$}}
\end{footnotesize}
\end{flushleft}

The third phase records that direction by replacing $T$'s head symbol \texttt{M} with an \texttt{A} or \texttt{B} (\texttt{A} in the example), performs some clean-up, replaces the symbol to be written, $S_{ij}$, stored in $s(t)$, with an \texttt{S}, and instead remembers $S_{ij}$. 

\begin{flushleft}
\begin{footnotesize}
\mbox{\texttt{...00000A000Y01SX0000001X0010110X0100011X0110100Y00...}} \newline
\mbox{\texttt{\ \ \ \ \ \ \ \ \ \ \ \ \ \ $\uparrow$}}
\end{footnotesize}
\end{flushleft}

The fourth and final phase performs the actual operations of $T$: it prints $S_{ij}$ (\texttt{1}) in the appropriate location on $T$'s tape, places an M to the left or right of that symbol depending on $D_{ij}$, and performs some final clean-up.

\begin{flushleft}
\begin{footnotesize}
\mbox{\texttt{...0000M1000Y01AX0000001X0010110X0100011X0110100Y00...}} \newline
\mbox{\texttt{\ \ \ \ \ \ \ \ \ \ \ \ \ \ $\uparrow$}}
\end{footnotesize}
\end{flushleft}

At this point, the first execution cycle is complete, and the second cycle begins.
\section{Exploiting the Universal Turing Machine}
\label{exploit}

Users of computer systems typically provide the input to, or argument of the computations, and are provided the results. From the point of view of computer security, it is typically undesirable to allow the user to subvert the functionality of the program performing the function. A malicious actor may, however, attempt to do so. A particularly serious security vulnerability is when it is possible for the end user to provide maliciously crafted data that effectively allows the execution of arbitrary code. In this section, we demonstrate that Marvin Minsky's universal Turing Machine suffers from an arbitrary code execution vulnerability. 

\subsection{Trust boundary}
There is one obvious trust boundary in a universal Turing machine, $U$: the initial string on the tape of the simulated Turing machine, $T$. That string corresponds to the user-provided data of an ordinary computer program. Because the potential users may be unknown to the developers and administrators of the computer and its programs, it is common to view this data as untrusted. In our explorations of the universal Turing machine, we will make the same assumption. Therefore, if it were possible to execute arbitrary code without manipulating the program of $T$, but only by providing crafted data on $T$'s simulated tape, that would constitute a vulnerability.

\subsection{Requirements on the machine description}
Nearly all possible machine descriptions of $T$ appear to be vulnerable to arbitrary code execution. We consider the case when the first executed quintuple is of the form $Q_0S_0Q_{01}S_{01}$\texttt{0}. The final symbol is thus fixed to a \texttt{0}, indicating that the direction of the head must thus not shift right in the first execution cycle. This is arguably in line with Minsky, p.138, \cite{minsky1967computation}: "We will assume that $T$ has a tape which is infinite only to the left [...]". Because exploitation occurs already in the first executed quintuple, additional quintuples will not affect the outcome.

\subsection{The exploit}
The following crafted input data will achieve arbitrary code execution by injecting a new Turing machine $I$, and coercing $U$ into simulating it:

\begin{center}
\mbox{\texttt{$\Delta$Y$q(i)$AX$Q_0S_0Q_{0x}S_{0x}D_{0x}$X$Q_1S_1Q_{1y}S_{1y}D_{1y}$...S}} \newline
\end{center}

where $\Delta$ represents the input data provided to the injected machine, $I$, $q(i)$ denotes the name of $I$'s current state, and $Q_xS_xQ_{xy}S_{xy}D_{xy}$ are the quintuples of $I$. All variable values need to be     coded as \texttt{B}s and \texttt{A}s instead of \texttt{1}s and \texttt{0}s. $q(i)=\mathtt{BA}$ if $S_{xy}=\mathtt{1}$ of the first executed quintuple and $q(i)=\mathtt{AA}$ if $S_{xy}=\mathtt{0}$.

\subsection{Example exploit}
We explain the exploitation mechanism by an example, where $T$'s machine description, $d_T$, consists of a simple program acting as a  binary counter, according to Figure \ref{fig-bc}, 

\begin{center}
\begin{footnotesize}
\mbox{$d_T=\mathtt{0000001X0010110X0100011X0110100}$} \newline
\end{footnotesize}
\end{center}

$U$'s initial tape is laid out as follows:

\begin{center}
\begin{footnotesize}
\mbox{\texttt{M000Y001X0000001X0010110X0100011X0110100Y00}} \newline
\end{footnotesize}
\end{center}

The injected machine, $I$, will aim to wipe the tape clean of user input, thus writing a \texttt{0} whenever encountering either a \texttt{0} or a \texttt{1}, and then shifting left. This can be accomplished with two quintuples:

\begin{center}
\begin{footnotesize}
\mbox{$d_I=\mathtt{0000000X0010000}$} \newline
\end{footnotesize}
\end{center}

According to the previous subsection, the crafted input will take the form

\begin{center}
\begin{footnotesize}
\mbox{\texttt{1111YBAAXAAAAAAAXAABAAAAS}} \newline
\end{footnotesize}
\end{center}

where \texttt{1111} is the data on which the injected machine, $I$ will operate, $q(i)=\mathtt{BA}$ is the injected machine state, $s(i)=\mathtt{A}$ is the currently scanned symbol, and \texttt{AAAAAAAXAABAAAA} encodes $I$'s machine description, $d_I$, thus representing the wiper program.

\subsubsection{First execution cycle}
At the start of execution, $U$'s tape has the following appearance:

\begin{flushleft}
\begin{footnotesize}
\mbox{\texttt{...001111YBAAXAAAAAAAXAABAAAASM000Y001X0000001X0010110X0...}} \newline
\mbox{\texttt{\ \ \ \ \ \ \ \ \ \ \ \ \ \ \ \ \ \ \ \ \ \ \ \ \ \ \ \ \ \ \ \ \ \ \ \ \ \ $\uparrow$}}
\end{footnotesize}
\end{flushleft}

with $U$'s head positioned on the \texttt{X} between $T$'s currently scanned symbol, $s(t)$, and machine description, $d_T$. The first three phases of $U$ follow the description in Section \ref{utm}, finding the identity of the quintuple, \texttt{0010110} stored in the machine condition, $q(t)s(t)=\mathtt{001}$, and replacing that machine condition with the action part of the identified quintuple, $Q_TS_T=\mathtt{011}$. 

\begin{flushleft}
\begin{footnotesize}
\mbox{\texttt{...001111YBAAXAAAAAAAXAABAAAASA000Y01SX0000001X0010110X0...}} \newline
\mbox{\texttt{\ \ \ \ \ \ \ \ \ \ \ \ \ \ \ \ \ \ \ \ \ \ \ \ \ \ \ \ \ \ \ \ \ \ \ \ $\uparrow$}}
\end{footnotesize}
\end{flushleft}

In the fourth phase, $U$ aims to perform the action on $T$'s tape as specified by the retrieved quintuple. It does write a \texttt{1} at the expected location, but then, however, the crafted input, consisting of \texttt{A}s and \texttt{B}s instead of the expected \texttt{0}s and \texttt{1}s, causes $U$'s head to shift far left into the user-provided data, placing the marker, \texttt{M}, representing $T$'s head, at an unexpected location.

\begin{flushleft}
\begin{footnotesize}
\mbox{\texttt{...00111MYBAAXAAAAAAAXAABAAAAB1000Y01SX0000001X0010110X0...}} \newline
\mbox{\texttt{\ \ \ \ \ \ \ \ \ \ \ \ \ \ \ \ \ \ \ \ \ \ \ \ \ \ \ \ $\uparrow$}}
\end{footnotesize}
\end{flushleft}

At the end of $U$'s first execution cycle, not only $T$'s, but also $U$ head comes to rest further to the left than expected. Importantly, $U$'s head is located to the left of the symbol \texttt{Y} indicating the end of $T$'s tape. 

\subsubsection{Second execution cycle}

Because $U$'s head is located in the attacker-controlled segment of the tape, in it's attempt to identify the next quintuple to execute, the first phase of $U$'s second execution cycle mistakenly refers to the injected machine condition, $q(i)s(i)=\mathtt{BAA}$. Looking for \texttt{0}s and \texttt{1}s rather than \texttt{A}s and \texttt{B}s, it won't find anything in the injected machine description, $d_I$. Instead, it encounters the first match, \texttt{100} at a rather random location, just before the \texttt{Y} representing the end of $T$'s tape. The initial \texttt{1} is the result of $T$'s first and successful print operation. The ensuing \texttt{00} are simply a part of a buffer between $T$'s tape and machine condition, $Q_TS_T$, as introduced by Minsky in \cite{minsky1967computation}. 

\begin{flushleft}
\begin{footnotesize}
\mbox{\texttt{...00111MY100XAAAAAAAXAABAAAABBAA0Y01SX0000001X0010110X0...}} \newline
\mbox{\texttt{\ \ \ \ \ \ \ \ \ \ \ \ \ \ $\uparrow$}}
\end{footnotesize}
\end{flushleft}

In the second phase, attempting to collect the action part of the identified quintuple, $U$ will find the four digits closest to the right of its head. While these were supposed to constitute the tail end of a quintuple, they are instead are pieces of the aforementioned buffer, of $T$'s machine condition, $q(t)s(t)$, and $T$'s first quintuple, jointly creating the string \texttt{0010}, which is thus interpreted as $Q_{ij}S_{ij}D_{ij}$. Furthermore, in the middle of the attempt to copy the first three digits to the injected machine condition, $q(i)s(i)$, $U$ slips back to the right of the \texttt{Y} indicating the start of $T$'s machine condition, $q(t)s(T)$. The end result is that the first digit is copied to $q(i)$ while the remaining part is copied to $q(t)s(t)$. At the end of this phase, the complete tape has the following layout:

\begin{flushleft}
\begin{footnotesize}
\mbox{\texttt{...00111MYA00XAAAAAAAXAABAAAABBAAAYAASXA000001X0010110X0...}} \newline
\mbox{\texttt{\ \ \ \ \ \ \ \ \ \ \ \ \ \ $\uparrow$}}
\end{footnotesize}
\end{flushleft}

The third phase reverts the \texttt{A}s and \texttt{B}s to \texttt{0}s and \texttt{1}s, and replaces $T$'s head, \texttt{M}, with a symbol indicating the direction of the next shift. $U$'s head is once again positioned far into the untrusted, user-provided data.

\begin{flushleft}
\begin{footnotesize}
\mbox{\texttt{...00111AY00SX0000000X001000011000Y00SX0000001X0010110X0...}} \newline
\mbox{\texttt{\ \ \ \ \ \ \ \ \ \ \ \ $\uparrow$}}
\end{footnotesize}
\end{flushleft}

In the fourth and final phase of the second execution cycle, $U$ shifts $T$'s head one step and records in $I$'s machine condition, $q(i)s(i)$, the symbol under $M$. 

\begin{flushleft}
\begin{footnotesize}
\mbox{\texttt{...0011M0Y00BX0000000X001000011000Y00SX0000001X0010110X0...}} \newline
\mbox{\texttt{\ \ \ \ \ \ \ \ \ \ \ $\uparrow$}}
\end{footnotesize}
\end{flushleft}

At this point, the compromise is complete, as the injected machine, $I$, is syntactically correct, and the head of $U$ is located in the injected machine condition, $q(i)s(i)$, rather than in the originally intended one, $q(t)s(t)$. The head will never again traverse the \texttt{Y} denoting the end of $T$'s tape, interpreting it instead as the end of $I$'s machine description, $d_I$. 

\subsubsection{Subsequent execution cycles}
The following execution cycles will faithfully execute the injected machine, $I$, wiping the contents of the inputs provided $I$. 

\begin{flushleft}
\begin{footnotesize}
\mbox{\texttt{...001M00Y00BX0000000X001000011000Y00SX0000001X0010110X0...}} \newline
\mbox{\texttt{...00M000Y00BX0000000X001000011000Y00SX0000001X0010110X0...}} \newline
\mbox{\texttt{...0M0000Y00AX0000000X001000011000Y00SX0000001X0010110X0...}} \newline
\mbox{\texttt{...M00000Y00AX0000000X001000011000Y00SX0000001X0010110X0...}} \newline
\end{footnotesize}
\end{flushleft}
\section{Mitigations}
\label{mitigations}

 It is possible to improve on the Minsky implementation in order to mitigate the presented vulnerability. As a first mitigation strategy, we propose to validate inputs. This could be performed by introducing a preprocessing phase validating that the simulated machine's tape only consists of the expected \texttt{0}s and \texttt{1}s. 
 
 Secondly, we can restrict the execution space by reducing the number of defined quintuples. To simplify the finite-state description captured by Figure \ref{fig-utm}, Minsky declares many quintuples implicitly: ``The most common quintuples, of the form (qi, sj, qi, sj, dij) are simply omitted [in the diagram].'' \cite{minsky1967computation}. This is convenient, because explicitly adding all necessary such quintuples to the diagram would require close to 70 arrows in addition to the 45 currently in the diagram. However, because the implicit definition creates approximately twice as many quintuples as the 70 required, this strategy allows the machine to accept many tape symbols that are not necessary for the proper functioning of the machine. Of the 70 unnecessary quintuples, close to 30 were required in order to allow the exploitation demonstrated in the previous section. This mitigation does require some effort, as the 70 required quintuples must be specified. 
 
 Thirdly, we could fortify the division between program and data. In Minsky's implementation, only 17 of the 184 quintuples defining $U$ are supposed to operate on $T$'s tape. By, for instance, using special symbols on $T$'s tape, and ensuring that only the privileged 17 quintuples read and write on that alphabet, additional barriers to exploitation would be established.

\section{Discussion}
\label{discussion}

We return to the puzzling prevalence of arbitrary code execution vulnerabilities. Are these the result of some fundamental property of computing, or are they rather the consequence of coincidental, unfortunate decisions during the development of those systems? 

It is interesting to note that, as was the case for Minsky's universal Turing machine, arbitrary code execution vulnerabilities can be accidentally introduced even in the simplest computer model. Minsky obviously attempted to design neither a secure nor a vulnerable system, but despite his indifference, he happened to design a vulnerable machine. That would suggest that vulnerability is a property that is not unlikely to arise in universal Turing machines. The volume of vulnerabilities discovered in computer systems in recent decades would further support such a proposition. Is it then the case that computers are intrinsically brittle - that they at their very core have a propensity to arbitrary code execution vulnerabilities?

Considering the exploitation of the universal Turing machine in the previous section, we may speculate about reasons for such a potential propensity. One suggested root cause of computer insecurity is \textit{complexity} \cite{schneier1999plea}. While there is surely truth to that statement, the insecurity of Minsky's minimally complex computing machine would appear to indicate a propensity to vulnerability even in the absence of complexity. 

A related theory points the finger at the human factor \cite{gonzalez2002framework}; poor decision-making by people is the cause of insecurity. But also this hypothesis insufficiently explains the problems of Minsky's universal Turing machine. One could possibly argue that Marvin Minsky suffered from a lack of security awareness, but that would not alone explain the demonstrated possibility for the user to achieve code execution. There is also something inherent to the machine that makes it not only theoretically possible, but oftentimes also actually the case.

Another proposed theory blames John von Neumann's \textit{stored program concept} \cite{von1993first} for the woes of arbitrary code execution; the fact that data and program in computers are stored on the same storage medium may allow an attacker to illegitimately modify the program rather than the intended data \cite{paul1973method}. Considering the universal Turing machine, this does indeed seem to have something to do with its vulnerability. The exploit demonstrated in the previous section is striking in the manner the machine head so unquestioningly saunters from program to data. This does intimate that the answer is at least partially related to the flimsy border between program and data. However, it can only be a partial answer to the question. While the co-location of program and data in the same memory unit might be a cause for the vulnerability of the universal Turing machine, of the general prevalence of buffer overflows (CWE-119) and use-after-free vulnerabilities (CWE-416), it is less obvious how it would constitute a root cause of SQL injection vulnerabilities (CWE-89) or operating system command injection vulnerabilities (CWE-78). Return-oriented programming \cite{checkoway2010return} is another argument against the stored-program hypothesis: in such an attack, the program is never modified. Instead the program \textit{flow} is deftly controlled by the attacker, cherry picking among assembly statements in the unmodified original program text to concoct attacker-controlled behavior. Return-oriented programming can almost always substitute modification of the actual program. Even if programs are located in a memory separate from data, they typically need to be modifiable by some means - if they are hard coded, the infinitely flexible universal Turing machine reduces to yet another specific Turing machine. And if the programs are indeed located in a separate but writeable memory (e.g. as in the Modified Harvard Architecture), then it would appear that the vulnerabilities of modifiable program code are back, e.g. as demonstrated in \cite{francillon2008code}.

A final theory that may be on the cusp of explaining computers' propensity to vulnerability is suggested by Bratus et al. in the ;login: article \textit{Exploit Programming} \cite{bratus2011exploit}, considering the \textit{weird} and oftentimes surprisingly potent \textit{machines} that may appear when unexpected, crafted input is provided to a computer program. A remarkable prevalence of such machines, is, I believe, at the heart of the problem.

\section{Conclusion}
\label{conclusion}

This paper presents the discovery of an arbitrary code execution vulnerability in Marvin Minsky's 1967 universal Turing machine implementation. By submitting crafted input data, an attacker can coerce the machine into executing arbitrary instructions. While this vulnerability has no real-world implications, we discuss whether it indicates an intrinsic propensity for arbitrary code execution vulnerabilities in computers in general.

\section*{Availability}

A Python program implementing and exploiting the Minsky Turing machine considered in this paper is available on GitHub at https://github.com/intrinsic-propensity/turing-machine.

\bibliographystyle{plain}
\bibliography{Turing}

\end{document}